\documentclass[12pt,notitlepage,aps,preprintnumbers,superscriptaddress,nofootinbib,aps,prd,floatfix]{revtex4-2}
\usepackage{graphicx}
\usepackage{ifpdf}
\ifpdf
  \DeclareGraphicsExtensions{.pdf,.png,.jpg}
\else
  \DeclareGraphicsExtensions{.eps,.ps}
\fi
\usepackage{dcolumn}
\usepackage{bm}
\usepackage{slashed}
\usepackage{physics}
\usepackage{cancel}
\usepackage{extarrows}
\usepackage{multirow,array}
\usepackage{amsmath,amssymb,mathtools}
\usepackage{subfigure,rotating}
\usepackage{enumitem}
\setlist{nolistsep}
\usepackage{xcolor}
\usepackage{url}
\usepackage{slashed}
\usepackage{soul}
\usepackage{siunitx}
\usepackage{import}
\usepackage{hyperref}
\definecolor{nicered}{rgb}{0.5,0.,0.}
\definecolor{nicegreen}{rgb}{0.,0.5,0.}
\definecolor{niceblue}{rgb}{0.,0.,0.5}
\hypersetup{colorlinks,citecolor=nicegreen,linkcolor=nicered,urlcolor=niceblue}
\usepackage{orcidlink}
\usepackage{braket}
\usepackage{colortbl}
\usepackage{gensymb}
\usepackage{comment}
\renewenvironment{eqnarray}{\begin{equation}\begin{aligned}}
{\end{aligned}\end{equation}}

\newcommand{\GeV}{\textrm{GeV}}

\newcommand{\vv}[1]{\bm{\mathbf{#1}}}
\newcommand{\ddp}{\partial}
\newcommand{\mat}[1]{\hat{\vv{#1}}}

\begin{document}
\preprint{}
\title{DGLAP evolution at N\texorpdfstring{$^3$}{3}LO with the  \texttt{Candia} algorithm}

\author{Casey Hampson \orcidlink{0009-0004-9584-4253}}
\email{champso1@students.kennesaw.edu}
\affiliation{Department of Physics, Kennesaw State University, Kennesaw, GA 30144, USA}

\author{Marco Guzzi \orcidlink{0000-0003-3430-2691}}
\email{mguzzi@kennesaw.edu}
\affiliation{Department of Physics, Kennesaw State University, Kennesaw, GA 30144, USA}

\date{\today}
\begin{abstract}
We present a generalization of the $x$-space \texttt{Candia} algorithm to next-to-next-to-next-to-leading order (N$^3$LO) accuracy in Quantum Chromodynamics (QCD) for solving the DGLAP evolution equations for unpolarized parton densities in the nucleon.
The algorithm is based on logarithmic expansions of the solution and can be extended to all orders in QCD. 
An expansion equivalent to the exact solution of the DGLAP equation at N$^3$LO is presented in the non-singlet sector. 
Results for approximate N$^3$LO PDFs, evolved using the most recent approximations to the N$^3$LO DGLAP splitting functions, 
are provided for benchmarking. The new version of the code, \texttt{Candia-v2}, 
is publicly available at \url{https://github.com/champso1/candia-v2}~\cite{candiav2}.

\end{abstract}

\maketitle
\tableofcontents

\newpage 
\section{Introduction}
The collinear substructure of the nucleon is described by parton distribution functions (PDFs), which map out the longitudinal momentum fractions of its inner constituent quarks and gluons in high-energy collisions. PDFs are essential components of factorization formulae in Quantum Chromodynamics (QCD), which allow for the perturbative calculation of hadronic cross sections. A precise and accurate determination of PDFs is a major endeavor~\cite{Hou:2019efy,Bailey:2020ooq,NNPDF:2021njg} and is necessary to advance the current precision frontier of cross-section calculations at hadron colliders. In the QCD framework of collinear factorization, PDFs depend on two variables: the parton longitudinal momentum fraction $x$ and the energy scale $Q$ of the reaction. 
The PDF $x$-dependence is in general extracted from global QCD analyses of experimental data for a wide variety of hadronic cross sections \cite{Hou:2019efy,Bailey:2020ooq,NNPDF:2021njg,NNPDF:2024nan,McGowan:2022nag,Ablat:2024muy}. Significant efforts are currently dedicated to extracting PDFs from lattice QCD, though their precision does not yet rival that of global analyses. Recent progress in lattice-QCD PDFs is documented in~\cite{Lin:2025hka,Constantinou:2020hdm,Lin:2017snn,Cichy:2019ebf,Ma:2017pxb,Radyushkin:2017cyf} and references therein. 
The energy behavior of the PDFs instead, is predicted by perturbative QCD (pQCD) and is described by the Dokshitzer-Gribov-Lipatov-Altarelli-Parisi (DGLAP)~\cite{Gribov:1972rt,Gribov:1972ri,Lipatov:1974qm,Dokshitzer:1977sg,Altarelli:1977zs} integro-differential renormalization group equations. The latter require knowledge of the $x$-dependent PDF functional form extracted from experimental data at an initial energy scale $Q_0$.

The QCD precision-phenomenology program at hadron colliders is rapidly moving towards new high-precision standards where fixed-order cross sections in pQCD are obtained beyond the next-to-next-to-leading order (NNLO) accuracy in the QCD strong coupling $\alpha_s$. Recently, several hard-scattering contributions for Higgs boson production and the Drell-Yan process have been calculated at fixed order (FO) in pQCD at next-to-next-to-next-to-leading order (N$^3$LO)~\cite{Anastasiou:2015vya,Mistlberger:2018etf,Duhr:2019kwi,Baglio:2022wzu,Dreyer:2016oyx,Chen:2019lzz,Duhr:2020sdp,Duhr:2021vwj,Dulat:2017prg,Dulat:2018bfe,Chen:2021isd,Billis:2021ecs,Camarda:2021ict,Chen:2021vtu,Chen:2022lwc,Caola:2022ayt}. Deep Inelastic Scattering (DIS) at N$^3$LO in the zero-mass approximation has been known for a long time~\cite{Vermaseren:2005qc,Moch:2004xu,Moch:2007rq,Moch:2008fj} while
three-loop single-mass heavy-flavor corrections to the DIS structure functions~\cite{Ablinger:2025awb} as well as N$^{3}$LO corrections to jet production in DIS~\cite{Gehrmann:2018odt,Currie:2018fgr} have been recently computed. PDF extraction has also been recently performed at approximate N$^3$LO (aN$^3$LO)~\cite{NNPDF:2024nan,McGowan:2022nag} in QCD and with the inclusion of QED corrections with a photon PDF at NLO accuracy~\cite{Cridge:2023ryv,Ball:2025xgq}. 

aN$^3$LO PDFs are currently determined using approximate forms of the QCD DGLAP kernels. 
The calculation of the four-loop splitting functions, which are required to extend DGLAP evolution to full N$^3$LO accuracy, is a formidable task and is currently in progress~\cite{Gracey:1994nn,Davies:2016jie,Moch:2017uml,Gehrmann:2023cqm,Falcioni:2023tzp,Gehrmann:2023iah,Kniehl:2025ttz}. Once completed, this will allow the PDFs to match the accuracy of current N$^3$LO hard-scattering calculations, enabling the consistent computation of hadron-level cross sections at N$^3$LO. 

For current phenomenological applications, approximations to the DGLAP kernels based on calculations in the small-$x$~\cite{Bonvini:2018xvt} and large-$x$~\cite{Davies:2022ofz} limits, as well as on a sufficiently large number of Mellin moments, are available~\cite{Moch:2021qrk,Falcioni:2023luc,Falcioni:2023vqq,Moch:2023tdj,Falcioni:2024xyt,Falcioni:2024qpd}. These approximations ensure that the residual uncertainty in parton evolution remains at or below the percent level in kinematic regions relevant for most LHC applications.

Other critical ingredients of DGLAP evolution at N$^3$LO accuracy include the four-loop 
$\beta$-function~\cite{vanRitbergen:1997va,Czakon:2004bu}, together with the corresponding threshold matching conditions~\cite{Chetyrkin:1997sg}, as well as the calculation of operator matrix elements (OMEs)~\cite{Witten:1975bh,Buza:1995ie,Buza:1996wv} which represent the transition from a massless
parton to a heavy quark and include the heavy-quark mass dependence. These OMEs are required to match PDFs in a variable flavor number scheme (VFNS) in the presence of heavy-quark masses, including both single-mass and two-mass contributions. This milestone three-loop calculation in pQCD has been accomplished only recently~\cite{Ablinger:2025nnq,Ablinger:2025joi,Ablinger:2024qxg,Ablinger:2024xtt,Ablinger:2023ahe,Bierenbaum:2022biv,Ablinger:2022wbb,Behring:2021asx,Ablinger:2020snj,Ablinger:2019etw,Ablinger:2018brx,Ablinger:2017xml,Ablinger:2017err,Ablinger:2014tla,Ablinger:2014uka,Behring:2014eya,transitionAqg,Ablinger:2012qm,Blumlein:2012vq,Blumlein:2011mi,Ablinger:2010ty,Bierenbaum:2009zt,Bierenbaum:2008yu,Bierenbaum:2007qe,Bierenbaum:2007dm}.

The Run-3 of the Large Hadron Collider (LHC) is expected to deliver approximately 300 fb$^{-1}$ of integrated luminosity
and the future high-luminosity (HL-LHC) program aims at delivering an integrated luminosity ten times larger at both the ATLAS and CMS experiments. 
Moreover, high-precision data are going to be collected by the Electron Ion Collider (EIC)~\cite{AbdulKhalek:2021gbh}.
Such a high-precision standard in the experiments requires theory predictions for the hadronic cross sections of comparable quality. Therefore, PDFs with improved uncertainties are needed, which must encode all the theoretical advances discussed above.

Parton evolution is crucial for the success of all these programs and past and recent literature contains multiple efforts devoted to the development of algorithms using different approaches to solve DGLAP equations~\cite{Furmanski:1981cw,Rossi:1983xz,DaLuzVieira:1990xk,Kumano:1992vd,Pascaud:1994vx,Miyama:1995bd,Pascaud:1996ci,Kosower:1997hg,Hirai:1997gb,Blumlein:1997em,Coriano:1998wj,Santorelli:1998yt,Ratcliffe:2000kp,Weinzierl:2002mv,Cafarella:2005zj,Jadach:2005bf,Guzzi:2006wx,Jadach:2007qa,Simonelli:2024vyh}. In the past three decades, part of these methods have been implemented in efficient computer codes, some of which are publicly available. Examples of these codes are \texttt{partonevolution}~\cite{Weinzierl:2002mv} (Mellin-transform method with an optimized contour), \texttt{Pegasus}~\cite{Vogt:2004ns} ($U$-matrix method in Mellin space), \texttt{Candia}~\cite{Cafarella:2008du,candiav1} (logarithmic recursion relations in the $x$-space), \texttt{Hoppet}~\cite{Salam:2008qg,Karlberg:2026kte} (Runge-Kutta method in the $x$-space), \texttt{QCDNum}~\cite{Botje:2010ay} (polynomial spline interpolation in the $x$-space), 
and more recently 
\texttt{Apfel}~\cite{Bertone:2013vaa,Bertone:2017gds} (Runge-Kutta method and higher-order interpolations in the $x$-space), \texttt{ChiliPDF}~\cite{Diehl:2021gvs} (global interpolation with Chebyshev polynomials in Mellin space),
\texttt{EKO}~\cite{Candido:2022tld} (evolution kernel operator method in Mellin space),
\texttt{uPDFevolv}~\cite{Hautmann:2014uua,Jung:2024uwc} (parton branching method). A recent benchmarking exercise of QCD evolution at aN$^3$LO is reported in Ref.~\cite{Cooper-Sarkar:2024crx}. 

With the recent advancement in the splitting function and OME calculations discussed above, the goal of this work is to extend the algorithm~\cite{Cafarella:2005zj,Guzzi:2006wx} implemented in the \texttt{Candia} computer code~\cite{Cafarella:2008du,candiav1} to N$^3$LO in QCD. The algorithm is based on logarithmic expansions of the DGLAP solution obtained by recursion relations evaluated directly in the $x$-space. It produces truncated solutions at arbitrary order in $\alpha_s$ for DGLAP equations with NLO, NNLO, and aN$^3$LO kernels.  
The strength of the algorithm lies in its ability to provide exact solutions to the non-truncated DGLAP equation in the non-singlet (NS) sector, expressed as a power series, with the added advantage of offering insight into the analytical structure of the solution. This series converges rapidly, after a relatively small number of iterations (${\cal O}(10)$). In this work, we present new recursion relations for the exact solutions in power series for the NS sector at N$^3$LO in QCD as well as the N$^3$LO-extended recursion relations for truncated solutions at arbitrary order for both the Singlet (S) and NS sectors. 
The mathematical equivalence between the \texttt{Candia} method in $x$-space and the $U$-matrix approach~\cite{Furmanski:1981cw} implemented in \texttt{Pegasus} in Mellin space~\cite{Vogt:2004ns} has been documented in previous work from one of the authors in Refs.~\cite{Cafarella:2005zj,Cafarella:2008du,Guzzi_2006} which we refer to throughout this work.
This study focuses on DGLAP QCD solutions and the generalization to QED with the inclusion of a photon PDF will be presented in future work. 
The \texttt{Candia} computer code which we release together with this manuscript, has been improved in several ways as compared to its first release~\cite{Cafarella:2008du}. It has been completely rewritten in \texttt{C++} using the \texttt{CMake} platform to improve performance, and various numerical optimizations have been implemented which dramatically improve the CPU turnaround time and memory usage. This new version is named \texttt{Candia-v2} and is available on the GitHub platform~\cite{candiav2}.

The manuscript is organized as follows. In Sec.~\ref{Notation}, we introduce the notation and various definitions used throughout this work. In Sec.~\ref{NS-solutions}, we present an exact solution in power series to the DGLAP equations in the NS sector while truncated solutions, extended to arbitrary perturbative order in QCD for both the S and NS sectors are presented in Sec.~\ref{S-NS-trunc-sol}.
The factorization and renormalization scale dependence is discussed in Sec.~\ref{scale-dep} while the threshold matching conditions for PDFs are in Sec.~\ref{threshold-cond}. In Sec.~\ref{results}, we show the N$^3$LO parton evolution results for the \texttt{Candia-v2} implementation obtained with the recent aN$^3$LO splitting functions and in Sec.~\ref{optmiz}, we discuss optimizations and improvements compared to the original 2008 release of \texttt{Candia}. We report our summary and conclusions in Sec.~\ref{conclusions}.

\section{DGLAP Equations: Notation and Definitions}
\label{Notation}

The DGLAP renormalization group equations (RGEs) are a direct consequence of QCD factorization which predict the energy scale evolution of the PDFs by requiring physical observables to be independent of the factorization scale $\mu_F$. They are written as

\begin{equation}
  \frac{\ddp f_{i/h}(x,\mu_F^2)}{\ddp \ln \mu_F^2} =\sum_{j} \mathcal{P}_{ij}(x,\alpha_s) \otimes f_{j/h}(x,\mu_F^2),
\label{eq:dglap}
\end{equation}

where $f_{i/h}(x,\mu_F^2)$ is the PDF representing the probability for parton $i$ (i.e., $q_i,{\bar q}_i, g$) to be emitted from hadron $h$ with longitudinal momentum fraction $x$ at scale $\mu_F^2$. The $\mathcal{P}_{ij}(x,\alpha_s)$ functions are the splitting functions (also known as DGLAP kernels) which represent the probability of finding a parton $i$ inside parton $j$. They are perturbatively calculable and their expansion
in $\alpha_s$ is given by

\begin{equation}
  \mathcal{P}(x,\alpha_s) = \sum_{n=0}^{\infty} \left( \frac{\alpha_s}{2\pi} \right)^{n+1} P^{(n)}(x).
\end{equation}

where the 1-loop $P^{(0)}(x)$ first appeared in Refs.~\cite{Gross:1973id,Politzer:1974fr,Altarelli:1977zs}, the 2-loop $P^{(1)}(x)$ is in Refs.~\cite{Floratos:1977au,Floratos:1978ny,Gonzalez-Arroyo:1979guc,Gonzalez-Arroyo:1979qht,Curci:1980uw,Furmanski:1980cm,Floratos:1981hs,Hamberg:1991qt}, the 3-loop $P^{(2)}(x)$ calculation is  in Refs.~\cite{Moch:2004pa,Vogt:2004mw}, and the calculation of $P^{(3)}(x)$ is in progress~\cite{Gracey:1994nn,Davies:2016jie,Moch:2017uml,Gehrmann:2023cqm,Falcioni:2023tzp,Gehrmann:2023iah,Kniehl:2025ttz}. The $\otimes$ symbol represents the convolution integral defined as
\begin{equation}
  [a \otimes b](x) \equiv \int_0^1 \frac{\dd y}{y} \; a(y) b\left( \frac{x}{y} \right).
\end{equation}

The running of the QCD strong coupling is governed by the renormalization group equations (RGEs)  
\begin{equation}
\frac{\ddp \alpha_s(\mu_R)}{\ddp \ln \mu_R^2} = \beta(\alpha_s) = -\sum_{n=0}^{\infty} \frac{\beta_n}{(4\pi)^{n+1}} \alpha_s^{n+2},
\label{eq:1-basics-betafn}
\end{equation}

where $\mu_R$ is the renormalization scale and the $\beta_n$ coefficients of the QCD $\beta$-function~\cite{Gross:1973id,Politzer:1974fr}  are known up to four loops~\cite{vanRitbergen:1997va,Czakon:2004bu}
\begin{align}
\beta_{0}  &=  \frac{11}{3}N_{C}-\frac{4}{3}T_{f} \\
\beta_{1}  &=  \frac{34}{3}N_{C}^{2}-\frac{20}{3}N_{C}T_{f}-4C_{F}T_{f}   
\\
\beta_{2}  &= 
\frac{2857}{54}N_{C}^{3}+2C_{F}^{2}T_{f}-\frac{205}{9}C_{F}N_{C}T_{f}-\frac{1415}{27}N_{C}^{2}T_{f}
+\frac{44}{9}C_{F}T_{f}^{2}+\frac{158}{27}N_{C}T_{f}^{2} 
\\
\beta_3 & =   \left( \frac{149753}{6} + 3564 \zeta_3 \right)
        - \left( \frac{1078361}{162} + \frac{6508}{27} \zeta_3 \right) n_f
  \nonumber \\ 
       &+ \left( \frac{50065}{162} + \frac{6472}{81} \zeta_3 \right) n_f^2
       +  \frac{1093}{729}  n_f^3\,,
\label{eq:beta3}
\end{align}

where $\beta_3$ is from Ref.~\cite{vanRitbergen:1997va} and

\begin{equation}
N_{C}=3,\qquad C_{F}=\frac{N_{C}^{2}-1}{2N_{C}}=\frac{4}{3},\qquad
T_{f}=T_{R}n_{f}=\frac{1}{2}n_{f}.
\end{equation}
$N_{C}$ is the number of colors, and $n_{f}$ is the number of
active flavors according to the condition $m_{q}\leq \mu_F$ for a given factorization scale
$\mu_F$, where $m_q$ is the mass of the $q$ quark.

Using Eq.~(\ref{eq:1-basics-betafn}), DGLAP equations can be rewritten as 

\begin{equation}
  \label{eq:dglap-main}
  \frac{\ddp f_{i/h}(x,Q^2)}{\ddp \alpha_s(Q^2)} = \sum_j \frac{\mathcal{P}_{ij}(x,Q^2)}{\beta(\alpha_s(Q^2))} \otimes f_{j/h}(x,Q^2),
\end{equation}

with the derivative in terms of $\alpha_s$ rather than $\ln Q^2$, where $Q^2$ is the evolution scale.

\subsection{Singlet and Non-Singlet DGLAP Sectors}
\label{S-NS-sectors}

For completeness and self-consistency, in this section we briefly summarize the notation and definitions, largely following the conventions of Refs.~\cite{Vogt:2004mw,Vogt:2004ns,Cafarella:2005zj,Cafarella:2008du}.

{\bf The Non-Singlet sector.} DGLAP equations are divided into two sets according to the $SU(n_f)$ flavor structure: the non-singlet (scalar) and singlet (matrix) equations where PDFs evolve independently. Flavor symmetry and the invariance under charge conjugation constrain the structure of the quark anti-quark splitting functions such that they can be written as
\begin{align}
  P_{q_iq_k} &= P_{\bar{q}_i\bar{q}_k} = \delta_{ik} P_{qq}^V + P_{qq}^S, \\
  P_{q_i\bar{q}_k} &= P_{\bar{q}_iq_k} = \delta_{ik}P_{q\bar{q}}^V + P_{q\bar{q}}^S\,,
\end{align}
where the $x$ and scale dependence has been omitted to simplify the notation, and where $P_{qq}^V$ and $P_{q\bar{q}}^V$ are the ``valence'' flavour-diagonal, and $P_{qq}^S$ and $P_{q\bar{q}}^S$ are the ``sea'' flavour-independent kernels. The flavor combinations
\begin{equation}
q_{i}^{(\pm)}=q_{i}\pm\overline{q}_{i}\,,
~~~~~~q^{(\pm)}=\sum_{i=1}^{n_{f}}q_{i}^{(\pm)},
\label{NS-S-comb}
\end{equation}
are used to construct the NS flavor asymmetries 
\begin{equation}
q_{\mathrm{NS},ik}^{(\pm)} = q_i^{(\pm)} -q_k^{(\pm)}
\end{equation}
which together with $q^{(-)}$, separately evolve with the kernels 
\begin{equation}
P_{\mathrm{NS}}^{(\pm)} = P_{qq}^V \pm P_{q\bar{q}}^V
~~~\textrm{and} ~~~
P_{\mathrm{NS}}^V = P_{qq}^V - P_{q\bar{q}}^V + n_f(P_{qq}^S - P_{q\bar{q}}^S) \equiv P_{\mathrm{NS}}^- + P_{\mathrm{NS}}^S\,,
\end{equation}
respectively in the NS sector. 

{\bf The Singlet sector.} The DGLAP equations in 
the singlet sector are matrix equations that couple the singlet-quark combination and gluon distributions
\begin{equation}
\frac{d}{d\ln \mu_F^2} \begin{pmatrix}q^{(+)} \\ g\end{pmatrix} = \begin{pmatrix}P_{qq} & P_{qg} \\ P_{gq} & P_{gg}\end{pmatrix} \otimes \begin{pmatrix}q^{(+)} \\ g\end{pmatrix},
\label{eq:dglap-singlet}
\end{equation}
where $q^{(+)}$ is defined in Eq.~(\ref{NS-S-comb}) and the quark-quark splitting function $P_{qq}$ is written as
\begin{equation}
P_{qq}=P_{NS}^{+}+n_{f}\left(P_{qq}^{S}+P_{q\bar{q}}^{S}\right)\equiv P_{NS}^{+}+P_{ps}\,,
\label{Pqq-expr}
\end{equation}
where $ps$ denotes ``pure-singlet'' terms. The splitting functions relative to gluon-quark and quark-gluon are $P_{qg}=n_{f}P_{q_{i}g}$ and $P_{gq}=P_{gq_{i}}$ respectively, expressed in terms of the flavor-independent splitting functions $P_{q_{i}g}=P_{\bar{q}_{i}g}$
and $P_{gq_{i}}=P_{g\bar{q}_{i}}$.
The reconstruction of the quark flavors at various perturbative orders proceeds as follows. 

{\bf Flavor reconstruction at N$^3$LO and NNLO.} 
From the singlet sector, we obtain the two independent $g$ and $q^{(+)}$ singlet distributions and to evolve the remaining $2n_{f}-1$ independent NS distributions, we select: $1)$ $q^{(-)}$ which evolves with $P_{NS}^{V}$; $2)$ $q_{NS,1i}^{(-)}=q_{1}^{(-)}-q_{i}^{(-)}$ (for 
$2\leq i\leq n_{f}$) which evolves with $P_{NS}^{-}$; $3)$ $q_{NS,1i}^{(+)}=q_{1}^{(+)}-q_{i}^{(+)}$ (for 
$2\leq i\leq n_{f}$) which evolves with $P_{NS}^{+}$. We use the relations 
\begin{equation}
q_{i}^{(\pm)}=\frac{1}{n_{f}}\left(q^{(\pm)}+\sum_{k=1,k\neq i}^{n_{f}}q_{NS,ik}^{(\pm)}\right)
\label{linNScomb}
\end{equation}
to reconstruct the quark flavors. 
The $q_{1}^{(-)}$ distribution is obtained by selecting $i=1$ in Eq.~(\ref{linNScomb}) and by
using the evolved NS distributions 1) and 2). 
The $q_{1}^{(+)}$ is obtained from
the evolved singlet $q^{(+)}$ and the NS distribution 3). 
Finally, for each flavor $i$ in the $2\leq i\leq n_{f}$ range we calculate $q_{i}^{(-)}$ and
$q_{i}^{(+)}$ from the NS distributions 2) and 3), and ultimately we obtain the individual flavors $q_{i}$ and $\bar{q}_{i}$.

{\bf Flavor reconstruction at NLO and LO.} At perturbative orders below NNLO, flavor reconstruction becomes easier because of the relation $P_{qq}^{S,(1)}=P_{q\bar{q}}^{S,(1)}$, which makes $P_{NS}^{V,(1)}=P_{NS}^{-,(1)}$.
Therefore, the NS $q^{(-)}$ and $q_{NS,ik}^{(-)}$ distributions evolve with
the same kernel, and so does $q_{i}^{(-)}$ for each flavor $i$. The basis for the remaining $2n_{f}-1$ NS distributions that is evolved at NLO is: 1) $q_{i}^{(-)}$ (for each $i\leq n_{f}$), which evolves with $P_{NS}^{-,(1)}$; 
2) $q_{NS,1i}^{(+)}=q_{1}^{(+)}-q_{i}^{(+)}$ (for each $i$ such that
$2\leq i\leq n_{f}$) which evolves with $P_{NS}^{+,(1)}$. The same basis is chosen at LO, where we also have the $P_{NS}^{+,(0)}=P_{NS}^{-,(0)}$ condition, 
since $P_{q\bar{q}}^{V,(0)}=0$.

\section{N$^3$LO Non-Singlet Exact Solutions in Power Series}
\label{NS-solutions}

In this section, we present the \texttt{Candia} algorithm for the exact DGLAP solutions in power series in the NS sector at N$^3$LO in QCD. The NLO and NNLO cases were presented and extensively discussed in Refs.~\cite{Cafarella:2005zj,Cafarella:2008du,Guzzi:2006wx}. We start by writing the non-truncated N$^3$LO scalar DGLAP equation for the generic NS combination $f_{NS,i}(x,\mu^2)$ at the scale $\mu^2$
\begin{equation}
\frac{\ddp{f_{NS,i}(x,\alpha_s)}}{\ddp{\alpha_s}} = \frac{\left( \frac{\alpha_s}{2\pi} \right)P^{(0)} + \left( \frac{\alpha_s}{2\pi} \right)^2 P^{(1)} + \left( \frac{\alpha_s}{2\pi} \right)^3 P^{(2)} + \left( \frac{\alpha_s}{2\pi} \right)^4 P^{(3)}}{- \frac{\alpha_s^2}{4\pi}\beta_0 - \frac{\alpha_s^3}{16\pi^2}\beta_1 - \frac{\alpha_s^4}{64\pi^3}\beta_2 - \frac{\alpha_s^5}{256\pi^4}\beta_3}\otimes f_{NS,i}(x,\alpha_s)\,,
\label{eq:n3lo-dglap}
\end{equation}

where the $\mu$-scale and $x$-dependence in $\alpha_s$ and in the splitting functions are  dropped for simplicity. The formal solution of Eq.~(\ref{eq:n3lo-dglap}) in $x$-space can be written as 

\begin{equation}
f^{\textrm{N}^3\textrm{LO}}(x,\mu^2) = e^{L~ a(x)\otimes} ~e^{M~ b(x)\otimes}~ e^{Q~ c(x)\otimes} ~ e^{W ~d(x)\otimes}~ f(x,\mu_0^2),
\label{formal-sol}
\end{equation}

where $\mu_0$ is the initial energy scale and $f(x,\mu_0^2)$ is defined by initial conditions. The functions $a(x), \dots, d(x)$ (functions of $P^{(0)}(x), \dots, P^{(3)}(x)$) as well as the coefficients $L,\dots,W$ (logarithms of functions of $\alpha_s(\mu)$ and $\alpha_s(\mu_0)\equiv \alpha_0$), must be determined. 
Eq.~(\ref{formal-sol}) is interpreted as the convolution series 

\begin{eqnarray}
e^{F P(x)\otimes} &\equiv& \sum_{n=0}^\infty \frac{F^n}{n!}\left(P(x)\otimes\right)^n
\end{eqnarray}

acting on a generic initial-condition function 
$\phi(x,\mu_0)$ as

\begin{eqnarray}
e^{F P(x)\otimes}\phi(x,\mu_0)&=&
\left( \delta(1-x)\otimes + F P(x)\otimes + \frac{1}{2!}F^2 P\otimes P\otimes
+ \dots\right)\phi(x,\mu_0) \nonumber \\
&=&
\phi(x,\mu_0) +  F \left(P \otimes \phi\right) + \frac{1}{2!}F^2 \left(P \otimes P \otimes \right)\phi(x,\mu_0)
\,+ \dots.\,,
\end{eqnarray}

where the coefficient $F$ depends on any other variable except $x$, and $P(x)$ is a generic splitting function. Associativity, distributivity and commutativity of the $\otimes$ product are obtained after mapping these products in Mellin space cfr.~\cite{Cafarella:2005zj}. For example,

\begin{eqnarray}
\mathcal{M}\left[\left(a\otimes b\right)\otimes c\right](N)=\mathcal{M}\left[ a\otimes \left(b\otimes c\right)\right](N) = A(N) B(N) C(N),
\end{eqnarray}

where $\mathcal{M}$ denotes the Mellin transform, $N$ is the Mellin-moment variable 
and $a(x)$, $b(x)$, and $c(x)$ are generic functions for which the $\otimes$ product is a regular function.  

The coefficients $L, \dots, W$ are obtained by analyzing the algebraic solution obtained in the Mellin space. The Mellin-space version of Eq.~(\ref{eq:n3lo-dglap}) after a partial fractioning on the polynomial ratio ${\cal P}(x,\alpha_s)/\beta(\alpha_s)$, is given by  

\begin{equation}
\frac{\ddp{f_{NS,i}(N,\alpha_s})}{\ddp{\alpha_s}} = \left[ \frac{R_0}{\alpha_s} + \frac{R'_1 + \alpha_s R'_2 + \alpha_s^2 R'_3}{64\pi^3\beta_0 + 16\pi^2\alpha_s \beta_1 + 4\pi \alpha_s^2 \beta_2 + \alpha_s^3 \beta_3} \right] f_{NS,i}(N,\alpha_s),
\label{eq:n3lo-dglap-apart}
\end{equation}

where
\begin{align}
R_0 &\equiv - \frac{2 P^{(0)}}{\beta_0}, 
~~~~R'_1 \equiv 32\pi^2 \beta_1/\beta_0 P^{(0)} - 64\pi^2P^{(1)}, \\
R'_2 &\equiv 8\pi \beta_2/\beta_0 P^{(0)} - 32\pi P^{(2)}, 
~~~
R'_3 \equiv 2\beta_3/\beta_0 P^{(0)} - 16 P^{(3)},
\end{align}
The exact algebraic solution of Eq.~(\ref{eq:n3lo-dglap-apart}) is obtained in Mellin space, where we integrate the right-hand side containing a cubic polynomial in the denominator with negative discriminant for each $n_f$ value. That is, there is one real root $r_1(n_f)$ and two complex conjugate roots $r_2(n_f)$ and $r_3(n_f)$, with $r_2(n_f)=r^{*}_3(n_f)$. Their numerical values for each $n_f$ are reported in Table~\ref{tbl:n3lo-roots}.
\begin{table}[ht]
\centering
\begin{tabular}{|c|c|c|}
\hline
$n_f$ & $r_1$ & $r_2 = r_3^*$ \\ \hline
1 & -0.96511 & 0.162930 - 0.95357i \\ \hline
2 & -1.03151 & 0.185237 - 1.02991i \\ \hline
3 & -1.11203 & 0.221422 - 1.13108i \\ \hline
4 & -1.20902 & 0.286759 - 1.27279i \\ \hline
5 & -1.32059 & 0.424770 - 1.48548i \\ \hline
6 & -1.42780 & 0.796497 - 1.81681i \\ \hline
\end{tabular}
\caption{Numerical values of the roots of the cubic polynomial in the denominator of Eq.~(\ref{eq:n3lo-dglap-apart}).}
\label{tbl:n3lo-roots}
\end{table}
The cubic polynomial can be factorized by leaving $r_1(n_f)$ in the linear term as
\begin{equation}
\beta_3[\alpha_s - r_1(n_f)][\alpha_s^2 + \bar{b}(n_f) \alpha_s + \bar{c}(n_f)],
\end{equation}
where $\bar{b} = -2\mathrm{Re}[r_2] = -2\mathrm{Re}[r_3]$, and $\bar{c} = |r_2|^2 = |r_3|^2$. 
After these substitutions, Eq.~(\ref{eq:n3lo-dglap-apart}) becomes   
\begin{equation}
\frac{\ddp{f_{NS,i}(N,\alpha_s)}}{\ddp{\alpha_s}} = \left[ \frac{R_0}{\alpha_s} + \frac{R'_1 + \alpha_s R'_2 + \alpha_s^2 R'_3}{\beta_3(\alpha_s - r_1)(\alpha_s^2 + \bar{b}\alpha_s + \bar{c})} \right]f_{NS,i}(N,\alpha_s)\,,
\label{partial-frac}
\end{equation}
whose N-space solution is given by
\begin{align}
f_{NS,i}(N,\alpha_s) = &\textrm{exp}\Bigg\{ A(N)\log \frac{\alpha_s}{\alpha_0}  + B(N)\log(\frac{\alpha_s^2 + \bar{b}\alpha_s + \bar{c}}{\alpha_0^2 + \bar{b}\alpha_0 + \bar{c}}) + C(N)\log(\frac{\alpha_s - r_1}{\alpha_0 - r_1}) 
\nonumber\\
&+ \frac{D(N)}{\sqrt{-\bar{b}^2 + \bar{c}}}\arctan\left(\frac{(\alpha_s-\alpha_0)\sqrt{-\bar{b}^2+\bar{c}}}{\alpha_s \bar{b} + \alpha_0(2\alpha_s + \bar{b}) + 2\bar{c}}\right)\Bigg\}f_{NS,i}(N,\alpha_0)\,, 
\label{eq:n3lo-full-sol}
\end{align}
where $-\bar{b}^2(n_f)+\bar{c}(n_f)>0$ for all $n_f$ values, and the $A(N),\dots,D(N)$ Mellin functions are
\begin{align}
A(N) &= R_0, ~~~\gamma \equiv r_1^2 + \bar{b} r_1 + \bar{c} \,,
\nonumber\\
B(N) &= \frac{1}{2\gamma}[-R'_1 -r_1 R'_2 + (\bar{c} + \bar{b} r_1)R'_3], 
\nonumber\\
C(N) &= \frac{1}{\gamma}(R'_1 + r_1 R'_2 + r_1^2 R'_3), 
\nonumber\\
D(N) &= -\frac{1}{\gamma}[(\bar{b} + 2r_1) R'_1 - (2\bar{c} + \bar{b} r_1) R'_2 + (\bar{b}\bar{c} + \bar{b}^2 r_1 - 2\bar{c} r_1) R'_3].
\label{eq:n3lo-resum-coeffs}
\end{align}
Eq.~(\ref{eq:n3lo-full-sol}) allows us to determine the coefficients $L,\dots,W$ in the formal N$^3$LO solution in Eq.~(\ref{formal-sol}) for the NS sector
\begin{align}
L &= \log \left(\frac{\alpha_s}{\alpha_0}\right), 
\\
M &= \log \left(\frac{\alpha_s^2 + \bar{b}\alpha_s + \bar{c}}{\alpha_0^2 + \bar{b}\alpha_0 + \bar{c}}\right), 
\\
Q &= \log \left(\frac{\alpha_s - r_1}{\alpha_0 - r_1}\right). 
\\
W &= \frac{1}{\sqrt{-\bar{b}^2+4\bar{c}}}\arctan\left( \frac{(\alpha_s - \alpha_0)\sqrt{-\bar{b}^2+4\bar{c}}}{2\alpha_s\alpha_0 + \bar{b}(\alpha_s+\alpha_0) + 2\bar{c}} \right),
\end{align}
while the $x$-space counterparts $a(x),\dots,d(x)$ of the N-space functions $A(N),\dots,D(N)$ are obtained from recursion relations generated by the following $x$-space ansatz   

\begin{small}
\begin{eqnarray}
f^{\mathrm{N}^3\mathrm{LO}}(x,\mu^2) = \left(\sum_{n=0}^\infty \frac{a(x)^n}{n!}L^n\right)_{\otimes}
\left(\sum_{m=0}^\infty \frac{b(x)^m}{m!}M^m\right)_{\otimes}
\left(\sum_{\ell=0}^\infty \frac{c(x)^\ell}{\ell!}Q^\ell\right)_{\otimes}
\left(\sum_{k=0}^\infty \frac{d(x)^k}{k!}W^k \right)_{\otimes} f(x,\mu_0^2).
\label{eq:n3lo-full-mellin-sol}
\end{eqnarray}
\end{small}

Combining the sums in Eq.~(\ref{eq:n3lo-full-mellin-sol}) with Cauchy products, the ansatz for the solution to N$^3$LO DGLAP equations in the NS sector becomes 
\begin{equation}
f^{\mathrm{N}^3\mathrm{LO}}(x,\mu^2) = \sum_{s=0}^\infty\sum_{t=0}^s\sum_{m=0}^t\sum_{n=0}^m \frac{D^s_{t,m,n}(x)}{n!(m-n)!(t-m)!(s-t)!}L^n M^{m-n}Q^{t-m}W^{s-t},
\label{eq:n3lo-ansatz}
\end{equation}
where $s=n+m+\ell+k$ and $t=n+m+\ell$, and  where we have introduced the coefficients
\begin{equation}
D_{t,m,n}^{s}(x)= a_{n}(x)\otimes
b_{m-n}(x)\otimes c_{t-m}(x) \otimes d_{s-t}(x) \otimes f(x,\mu_0^2)
\end{equation}
to make the notation compact.

The N$^3$LO recursion relations to determine $D^s_{t,m,n}$ coefficients are obtained by substituting the ansatz in Eq.~(\ref{eq:n3lo-ansatz}) into Eq.~(\ref{partial-frac}) and equating equal powers of $\alpha_s$ on both sides. This procedure produces the four recursion relations below
\begin{align}
D^s_{t,m,n} &= Z_{11} \otimes D^{s-1}_{t-1,m-1,n-1}, \label{eq:n3lo-resum-1}\tag{\theequation{a}}\\
D^s_{t,m,n} &= Z_{21}\otimes D^s_{t,m,n+1} + Z_{22} \otimes D^{s-1}_{t-1,m-1,n}, 
\label{eq:n3lo-resum-2}\tag{\theequation{b}}\\
D^s_{t,m,n} &= Z_{31} \otimes D^s_{t,m+1,n+1} + Z_{32} \otimes D^{s-1}_{t-1,m,n}, 
\label{eq:n3lo-resum-3}\tag{\theequation{c}}\\
D^s_{t,m,n} &= Z_{41} \otimes D^s_{t+1,m+1,n+1} + Z_{42} \otimes D^s_{t+1,m+1,n} + Z_{43} \otimes D^{s-1}_{t,m,n}.\label{eq:n3lo-resum-4}\tag{\theequation{d}}
\end{align}
where the coefficients $Z_{ij}$ that include the dependence on the splitting functions and values of the roots of the cubic polynomial in the non-truncated ratio ${\cal P}(x,\alpha_s)/\beta(\alpha_s)$, are defined as
\begin{align}
Z_{11}(x) &= R_0(x) = -\frac{2}{\beta_0} P^{(0)}(x),  
\nonumber\\
Z_{21} &= \frac{1}{2\beta_3\gamma}\left[ 16\pi^2\beta_1 + 4\pi r_1\beta_2 - (\bar{c} + \bar{b} r_1)\beta_3 \right], 
\nonumber\\
Z_{22}(x) &= \frac{1}{\beta_3\gamma}\left[32\pi^2P^{(1)}(x) + 16\pi r_1 P^{(2)}(x) - 8(\bar{c}+\bar{b}r_1)P^{(3)} (x)\right], 
\nonumber\\
Z_{31} &=\frac{1}{\beta_3\gamma} \left[(-16\pi^2\beta_1 - 4\pi r_1\beta_2 - r_1^2\beta_3)\right], 
\nonumber\\
Z_{32}(x) &= \frac{1}{\beta_3\gamma} \left[- 64\pi^2 P^{(1)}(x) - 32\pi r_1 P^{(2)}(x) - 16 r_1^2 P^{(3)}(x)\right], 
\nonumber\\
Z_{41} &= -2 \bar{b},  
\nonumber\\
Z_{42} &= \frac{1}{\beta_3\gamma}\left[ 32\pi^2(\bar{b}+r_1)\beta_1  -8\pi \bar{c}\beta_2 - 2\bar{c}r_1\beta_3 \right],  
\nonumber\\
Z_{43}(x)& = \frac{1}{\beta_3\gamma}\left[ 128\pi^2(\bar{b}+r_1) P^{(1)}(x) - 64\pi \bar{c} P^{(2)}(x) - 32 \bar{c} r_1 P^{(3)}(x) \right]\,.
\label{Zij-def}
\end{align}
Note that not all the $Z_{ij}$ have a dependence on $x$, and convolution products in those cases reduce to algebraic multiplications.
The N$^3$LO exact NS solution in power series is then fully specified once all $D^{s}_{t,m,n}$ coefficients are determined. In fact, the recursion relations 
in Eqns.~(\ref{eq:n3lo-resum-1})--(\ref{eq:n3lo-resum-4}) can be solved enabling us to compute all the coefficients $D^{s}_{t,m,n}$ up to a chosen $s$ starting from the initial condition $f(x,\mu_0^2) = D^{0}_{0,0,0}$:  

\begin{eqnarray}
D^s_{t,m,n} = Z_{11}^n\otimes  
&\left[ Z_{11} Z_{21} + Z_{22} \right]^{m-n}\otimes \left[ Z_{11} Z_{31} + Z_{32} \right]^{t-m} \otimes
\\
&\left[ Z_{11} Z_{41} + Z_{42} (Z_{11} Z_{21} + Z_{22}) + Z_{43} \right]^{s-t}\otimes D^0_{0,0,0}.
\label{eq:n3lo-resummed-recrels}
\end{eqnarray}

It may seem that any of the recursion relations in Eqns.~(\ref{eq:n3lo-resum-1}-\ref{eq:n3lo-resum-4}) could in principle be used to determine a specific coefficient $D^{s}_{t,m,n}$. However, each recursion relation requires that particular other coefficients are known from previous steps, with the constraint that the initial condition $D^{0}_{0,0,0}$ is the only starting point for the evaluation. Therefore, there is only one procedure that can be followed for a given value of the outermost coefficient $s$ to determine all of the others:

\begin{enumerate}
\item All $D^s_{t,m,n}$ with $n\neq0$ are computed using Eq.~(\ref{eq:n3lo-resum-1}),
\item All $D^s_{s,s,0}$ are computed using Eq.~(\ref{eq:n3lo-resum-2}),
\item All $D^s_{s,m,0}$ are computed using Eq.~(\ref{eq:n3lo-resum-3}) where $m \neq s$, with decreasing $m$,
\item All $D^s_{t,m,0}$ are computed using Eq.~(\ref{eq:n3lo-resum-4}) where $t \neq s$, with decreasing $t$ and $m$.
\newline
\end{enumerate}

Finally, with the above definitions and procedure to determine $D^s_{t,m,n}$ in Eq.~(\ref{eq:n3lo-resummed-recrels}), the explicit $x$-space expression for the N$^3$LO solution to DGLAP equations in the NS sector is 
\begin{equation}
f^{N^3LO}(x,\mu^2) =  e^{\left\{  Z_{11} Z_{21} + Z_{22}\right\}_\otimes} e^{\left\{  Z_{11} Z_{31} + Z_{32}\right\}_\otimes} 
e^{\left\{Z_{11} Z_{41} + Z_{42} (Z_{11} Z_{21} + Z_{22}) + Z_{43}   \right\}_\otimes}
e^{R_0\otimes} D^0_{0,0,0}(x),
\label{n3lo-NSsolution}
\end{equation}
where the last factor with $R_0$ on the right-hand side (RHS), represents the LO contribution. It is easy to note that the expressions obtained by expanding the $Z_{ij}$ coefficients as in Eqns.~(\ref{Zij-def}) exactly correspond to the definitions in Eqns.~\eqref{eq:n3lo-resum-coeffs}.

\section{Truncated solutions for the DGLAP Singlet and Non-Singlet sectors at all orders in QCD}
\label{S-NS-trunc-sol}

In this section, we present recursion relations for the DGLAP truncated solutions expanded at arbitrary order in pQCD for both the singlet and NS sectors. The proof of existence of a valid logarithmic ansatz in $x$-space that reproduces such truncated solutions including the $P^{(1)}(x)$ and $P^{(2)}(x)$ splitting functions in NLO and NNLO DGLAP respectively, and proof of their equivalence to the solutions obtained with the $U$-matrix approach in Mellin space, has been given in Refs.~\cite{Cafarella:2005zj,Cafarella:2008du,Guzzi:2006wx}. Here we present the new recursion relations at arbitrary orders including $P^{(3)}(x)$. 

An expansion of the ${\cal \hat P}(x,\alpha_s)/\beta(\alpha_s)$ ratio in the DGLAP singlet equation at order ${\cal O}(\alpha_s^k)$ gives 
\begin{eqnarray}
\frac{\partial{{\bf f}(x,\alpha_s)}}{\partial\alpha_s}=
\frac{1}{\alpha_s}\left[{\bf R}_0+\alpha_s {\bf R}_1 +\alpha_s^2
{\bf R}_2+\alpha_s^3
{\bf R}_3 + \dots + \alpha_s^k {\bf R}_k \right]\otimes {\bf f}(x,\alpha_s),
\label{N3LOsinglet}
\end{eqnarray}
where bold-face characters are used for the vector notation. The ${\bf R}_k$ matrix operators are defined order by order as
\begin{eqnarray}
{\bf R}_0(x)&=-\frac{2}{\beta_0}{\bf P}^{(0)}(x),
~~~~
{\bf R}_1(x)=-\frac{1}{\pi}\left(\frac{\beta_1}{4\beta_0}{\bf R}_0(x) +
\frac{{\bf P}^{(1)}(x)}{\beta_0}\right),
\\
{\bf R}_2(x)&=-\frac{1}{\pi}\left(\frac{{\bf P}^{(2)}(x)}{2\pi\beta_0}
+\frac{{\bf R}_1(x) \beta_1}{4\beta_0}+\frac{{\bf R}_0(x) \beta_2}{16\pi\beta_0}\right), 
\nonumber\\
{\bf R}_3(x) &= -\frac{1}{\pi}\left(\frac{{\bf P}^{(3)}(x)}{4\pi^2\beta_0} +\frac{{\bf R}^{(2)}(x)\beta_1}{4\beta_0} + \frac{{\bf R}^{(1)}(x)\beta_2}{16\pi\beta_0}+\frac{{\bf R}^{(0)}(x)\beta_3}{64\pi^2\beta_0}\right),
\dots
\end{eqnarray}

The logarithmic ansatz for the truncated solution in the $x$-space that matches the algebraic solution in Mellin space at the arbitrary order $\kappa$ in $\alpha_s$ is given by
\begin{equation}
\vv{f}(x,\alpha_s) = \sum_{n=0}^\infty \left\{ \left[ \sum_{i=0}^\kappa \alpha_s^i \frac{\vv{S}^i_n(x)}{n!} \right] \ln^n \frac{\alpha_s}{\alpha_0} \right\},
\label{log-ansatz}
\end{equation}
where $\kappa\geq k$ represent the truncation index that defines the $\alpha_s$ accuracy of the singlet solution or the NS one, in the case of scalar equations. The $\vv{S}_n^i(x)$ coefficients  are determined by $x$-space recursion relations obtained by substituting Eq.~(\ref{log-ansatz}) in Eq.~(\ref{N3LOsinglet}) and equating each power of $\alpha_s$ on both sides of the equation. For consistency, here we report the recursion relations obtained at arbitrary order in $\alpha_s$ for the LO DGLAP case where only ${\bf P}^{(0)}$ appears, at NLO where ${\bf P}^{(0)}$ and ${\bf P}^{(1)}$ only appear, at NNLO with ${\bf P}^{(0)}$, ${\bf P}^{(1)}$ and ${\bf P}^{(2)}$ only, and finally at N$^3$LO with ${\bf P}^{(0)}$, ${\bf P}^{(1)}$, ${\bf P}^{(2)}$, and ${\bf P}^{(3)}$.

At LO (first power in $\alpha_s$), the solution can be given in closed form through the recursion relation below
\begin{equation}
\label{eq:lo-singlet-sol}
\vv{S}^0_{n+1} = - \frac{2}{\beta_0} \left[ \mat{P}^{(0)} \otimes \vv{S}^0_n \right](x).
\end{equation}

At NLO, where the $\mat{P}^{(0)}$ and $\mat{P}^{(1)}$ kernels appear, the $\bf S$ coefficients relative to the $i$-th ($i = 1,\dots,\kappa$) truncated solution are obtained through the recursion relation below  

\begin{align}
\vv{S}^i_{n+1}(x) = &-\frac{\beta_1}{4\pi\beta_0}\vv{S}^{i-1}_{n+1}(x) - i\vv{S}^i_n(x) - (i-1)\frac{\beta_1}{4\pi\beta_0}\vv{S}^{i-1}_n(x) \nonumber\\
&-\frac{2}{\beta_0}[\vv{P}^{(0)} \otimes \vv{S}^i_n](x) - \frac{1}{\pi\beta_0}[\vv{P}^{(1)} \otimes \vv{S}^{i-1}_n](x)\,,
\label{eq:nlo-singlet-i-sol}
\end{align}

while at NNLO, with ${\bf P}^{(0)}$, ${\bf P}^{(1)}$ and ${\bf P}^{(2)}$ only, we have 

\begin{align}
\vv{S}_{n+1}^i(x) = & -\frac{\beta_1}{4\pi\beta_0}\vv{S}_{n+1}^{i-1}(x)-\frac{\beta_2}{16\pi^2\beta_0}\vv{S}_{n+1} ^{i-2}(x) \nonumber\\
& -i\vv{S}_n^i(x)-(i-1)\frac{\beta_1}{4\pi\beta_0}\vv{S}_n^{i-1}(x)-(i-2)\frac{\beta_2}{ 16\pi^2\beta_0}\vv{S}_n^{i-2}(x) \nonumber\\
& -\frac{2}{\beta_0}[\vv{P}^{(0)}\otimes\vv{S}_n^i](x)
-\frac{1}{\pi\beta_0}[\vv{P}^{(1)}\otimes\vv{S}_n^{i-1}](x)
-\frac{1}{2\pi^2\beta_0}[\vv{P}^{(2)}\otimes\vv{S}_n^{i-2}](x)
\label{eq:nnlo-singlet-sol}\,.
\end{align}

At N$^3$LO, the recursion relation for the ${\bf S}$ coefficients is given by

\begin{align}
  \vv{S}_{n+1}^i(x) = &-\frac{\beta_1}{4\pi\beta_0}\vv{S}_{n+1}^{i-1}(x)
-\frac{\beta_2}{16\pi^2\beta_0}\vv{S}_{n+1} ^{i-2}(x)
-\frac{\beta_3}{64\pi^3\beta_0}\vv{S}^{i-3}_{n+1}(x) \nonumber\\
&-i\vv{S}_n^i(x)-(i-1)\frac{\beta_1}{4\pi\beta_0}\vv{S}_n^{i-1}(x)
-(i-2)\frac{\beta_2}{16\pi^2\beta_0}\vv{S}_n^{i-2}(x) 
-(i-3)\frac{\beta_3}{64\pi^3\beta_0}\vv{S}_n^{i-3}(x) \nonumber\\
&-\frac{2}{\beta_0}[\vv{P}^{(0)}\otimes\vv{S}_n^i](x)
-\frac{1}{\pi\beta_0}[\vv{P}^{(1)}\otimes\vv{S}_n^{i-1}](x)
-\frac{1}{2\pi^2\beta_0}[\vv{P}^{(2)}\otimes\vv{S}_n^{i-2}](x) \nonumber\\
&-\frac{1}{4\pi^3\beta_0}[\vv{P}^{(3)}\otimes\vv{S}_n^{i-3}](x).
\label{eq:singlet-n3lo}
\end{align}
These recursion relations generate truncated solutions at arbitrary order $\kappa$ that are equivalent those obtained using the $U$-matrix method~\cite{Blumlein:1997em,Vogt:2004ns,Cafarella:2005zj}.

\section{Factorization/Renormalization Scale Dependence}
\label{scale-dep}

The dependence of DGLAP evolutions on the factorization $\mu_F$ and renormalization $\mu_R$ scales treated as independent parameters, is acquired through the running of $\alpha_s$ in the splitting functions. In fact, expanding $\alpha_s(\mu_R^2)$ in terms of $\alpha_s(\mu_F^2)$, we obtain 
\begin{equation}
\alpha_s(\mu_R^2)=\alpha_s(\mu_F^2)-\left[-\alpha_s^2(\mu_F^2)\frac{\beta_0
L}{4\pi}
+\frac{\alpha_s^3(\mu_F^2)}{(4\pi)^2}(-\beta_0^2 L^2-\beta_1
L)\right] + \dots\,,
\end{equation}
where the dots denote higher-order terms and $L = \log(\mu_F^2/\mu_R^2)$. The $\mu_F$ and $\mu_R$ dependence explicitly appears in the splitting functions in terms of $L$ and we have 
\begin{equation}
\mathcal{P}(x,\mu_F^2,\mu_R^2) = \sum_{n=0}^{\infty} \left( \frac{\alpha_s(\mu_F)}{2\pi} \right)^{n+1} P^{(n)}(x, \mu_F^2/\mu_R^2),
\label{muR-muF-splitting}
\end{equation}
where
\begin{align}
P^{(0)} \left( x,\frac{\mu_F^2}{\mu_R^2} \right) =& P^{(0)}(x), 
\\
P^{(1)} \left( x,\frac{\mu_F^2}{\mu_R^2} \right) =& P^{(1)}(x) - \frac{\beta_0}{2}P^{(0)}(x)L,
\\
P^{(2)} \left( x,\frac{\mu_F^2}{\mu_R^2} \right) =& P^{(2)}(x) - \beta_0LP^{(1)}(x) + \frac{1}{4}(\beta_0^2L^2 - \beta_1L)P^{(0)}(x),
\\
P^{(3)} \left( x,\frac{\mu_F^2}{\mu_R^2} \right) =&  P^{(3)}(x) - \frac{3}{2}\beta_0LP^{(2)}(x) + \left( \frac{3}{4}\beta_0^2L^2 - \frac{1}{2}\beta_1L \right)P^{(1)}(x) \nonumber\\
&+ \frac{1}{8} \left( -\beta_0^3L^3 + \frac{5}{2}\beta_0\beta_1L^2 - \beta_2L \right)P^{(0)}(x),
\end{align}
and we have 
\begin{eqnarray}
  \frac{\ddp{f(x,\mu_F^2,\mu_R^2)}}{\ddp{\ln \mu_F^2}} = \mathcal{P}(x,\mu_F^2,\mu_R^2) \otimes f(x,\mu_F^2,\mu_R^2).
\end{eqnarray}
All parton indices in the equations above have been suppressed for simplicity. 

\section{Heavy-Quark matching conditions}
\label{threshold-cond}
In parton evolution, the number of active flavors changes as the energy scale $\mu$ crosses heavy–quark mass thresholds $m_h$ and collinear logarithms $\log(m_{h}^2/\mu^2)$ are resummed into PDFs. This allows for a transition from a theory with $n_f$ light flavors to one with $n_f+1$ light flavors at NNLO and beyond within a Variable Flavor Number Scheme (VFNS). Alternatively, the evolution may proceed with a fixed number of flavors in a Fixed Flavor Number Scheme (FFNS) regardless of which quark-mass threshold is crossed. The FFNS and VFNS options are available in the \texttt{Candia} evolution code. 

The decoupling at the Lagrangian level implies that Green functions of light fields in the $(n_f+1)$-flavor theory can be matched onto an effective $n_f$-flavor theory via decoupling constants. In the $\overline{\textrm{MS}}$ scheme, this is encoded in a decoupling relation for $\alpha_s$ and decoupling relations for local twist-2 operators whose matrix elements define the PDFs.

The matching condition for $\alpha_s$ up to N$^m$LO is given by~\cite{Chetyrkin:1997sg,vanRitbergen:1997va}

\begin{equation}
\alpha_s^{(n_f+1)}(k_r m_h^2) = \alpha_s^{(n_f)}(k_rm_h^2) + \sum_{n=1}^m \left( \alpha_s^{(n_f)}(k_rm_h^2) \right)^{n+1} \sum_{l=0}^n c_{n,l} \ln k_r,
\end{equation}

where $k_r$ is the ratio of the renormalization and factorization scales, $m_h^2$ is the mass of a heavy quark $h$ where $h=c, \, b, \, t$, and the $c_{n,l}$ coefficients are known up to four loops~\cite{Chetyrkin:1997sg,vanRitbergen:1997va}. 

The matching conditions for the light-quark PDFs are given as~\cite{Buza:1995ie,Buza:1996wv}
\begin{align}
l_i^{(n_f+1)}(x) + \bar{l}_i^{(n_f+1)}(x) &= l_i^{(n_f)}(x) + \bar{l}_i^{(n_f)}(x) \nonumber \\
&+ \left[ A_{qq,h}^{\mathrm{NS},+} \otimes \left( l_i^{(n_f)}(x) + \bar{l}_i^{(n_f)}(x) \right) \right](x) \nonumber \\
&+ \frac{1}{n_f} \Big\{ \left[A_{qq,h}^{\mathrm{PS}} \otimes q^{+,(n_f)} \right](x) + \left[ A_{qg,h}^{\mathrm{S}} \otimes g^{(n_f)} \right](x) \Big\}, \\
l_i^{(n_f+1)}(x) - \bar{l}_i^{(n_f+1)}(x) &= l_i^{(n_f)}(x) - \bar{l}_i^{(n_f)}(x) \nonumber \\
&+ \left[ A_{qq,h}^{\mathrm{NS},-} \otimes (l_i^{(n_f)}(x) - \bar{l}_i^{(n_f)}(x)) \right](x)\,,
\end{align}
for the gluon PDF we have
\begin{equation}
g^{(n_f+1)}(x) = g^{(n_f)}(x) + \left[ A_{gq,h}^{\mathrm{S}} \otimes q^{+,(n_f)} \right](x) + [A_{gg,h}^{\mathrm{S}} \otimes g^{(n_f)}](x),
\end{equation}
and for the heavy-quark PDFs, we have   
\begin{equation}
q_h^{(n_f+1)}(x) + \bar{q}^{(n_f+1)}(x) = [A_{hq}^{\mathrm{S}} \otimes q^{+,(n_f)}](x) + [A_{hg}^{\mathrm{S}} \otimes g^{(n_f)}](x).
\end{equation}
Recently, the heavy-flavor asymmetry has been computed in Ref.~\cite{Behring:2025avs} at N$^3$LO, whose behavior is controlled by the pure-singlet OME $A_{hq}^{\mathrm{PS},\mathrm{s},(3)}$. This gives the following new relation for the heavy-flavor difference PDF:
\begin{equation}
  q_h^{(n_f+1)}(x) - \bar{q}_h^{(n_f+1)} = [A_{hq}^{\mathrm{PS},\mathrm{s}} \otimes q^{-,(n_f)}](x).
\end{equation}
The OMEs have the perturbative expansions given below  
\begin{align}
A^{\textrm{NS},\pm}_{qq,h}(x) &= \left(\frac{\alpha_s(m_h^2)}{4\pi}\right)^2 A^{\textrm{NS},\pm,(2)}_{qq,h}(x) + 
\left(\frac{\alpha_s(m_h^2)}{4\pi}\right)^3 A^{\textrm{NS},\pm,(3)}_{qq,h}(x) \\
A^{\textrm{S}}_{gk,h}(x) &= \left(\frac{\alpha_s(m_h^2)}{4\pi}\right)^2 A^{\textrm{S},(2)}_{gk,h}(x) + 
\left(\frac{\alpha_s(m_h^2)}{4\pi}\right)^3 A^{\textrm{S},(3)}_{gk,h}(x),~~~~ k = q,g \\
A^{\textrm{S}}_{hk}(x) &= \left(\frac{\alpha_s(m_h^2)}{4\pi}\right)^2 A^{\textrm{S}(2)}_{hk}(x) + 
\left(\frac{\alpha_s(m_h^2)}{4\pi}\right)^3 A^{\textrm{S},(3)}_{hk}(x),~~~~ k = q,g \\
A^{\textrm{PS}}_{qq,h}(x) &= \left(\frac{\alpha_s(m_h^2)}{4\pi}\right)^3 A^{\textrm{PS},(3)}_{qq,h}(x)\,, \\
A^{\textrm{S}}_{qg,h}(x) &= \left(\frac{\alpha_s(m_h^2)}{4\pi}\right)^3 A^{\textrm{S},(3)}_{qg,h}(x)\,, \\
A_{hq}^{\mathrm{PS},\mathrm{s}}(x) &= \left(\frac{\alpha_s(m_h^2)}{4\pi}\right)^3 A_{hq}^{\mathrm{PS},\mathrm{s},(3)}(x)\,,
\end{align}
where the 2-loop OMEs $A^{(2)}_{ij}$ have been calculated in Refs.~\cite{Buza:1995ie,Buza:1996wv} while the most recent 3-loop calculations are presented in Refs.~\cite{Ablinger:2025nnq,Ablinger:2025joi,Ablinger:2024qxg,Ablinger:2024xtt,Ablinger:2023ahe,Bierenbaum:2022biv,Ablinger:2022wbb,Behring:2021asx,Ablinger:2020snj,Ablinger:2019etw,Ablinger:2018brx,Ablinger:2017xml,Ablinger:2017err,Ablinger:2014tla,Ablinger:2014uka,Behring:2014eya,transitionAqg,Ablinger:2012qm,Blumlein:2012vq,Blumlein:2011mi,Ablinger:2010ty,Bierenbaum:2009zt,Bierenbaum:2008yu,Bierenbaum:2007qe,Bierenbaum:2007dm,Behring:2025avs}.

\section{Results}
\label{results}

In this section, we present the numerical results from the N$^3$LO algorithm implemented in \texttt{Candia-v2} for solutions to the DGLAP evolution equations with aN$^3$LO splitting functions.
To facilitate benchmarking, we utilize the initial conditions given in Sec.~4.4 of Ref.~\cite{Dittmar:2005ed} at the initial scale of $\mu^2_0=2$ GeV$^2$ and evolve the PDFs in the VFNS up to $\mu_F^2 = 10^4$ GeV$^2$. The values of the heavy-quark masses in the pole-mass approximation are set as $m_c=\mu_0$, $m_b=4.5$ GeV, and $m_t=175$ GeV. For the aN$^3$LO splitting functions, we use the approximations in Refs.~\cite{Davies:2022ofz,Moch:2021qrk,Falcioni:2023luc,Falcioni:2023vqq,Moch:2023tdj,Falcioni:2024xyt,Falcioni:2024qpd} while for those at NNLO, we use the approximations in Refs.~\cite{Moch:2004pa,Vogt:2004mw} for a fast evaluation. The 3-loop OMEs for heavy-quark threshold conditions beyond NNLO with single mass are included as in Refs.~\cite{Ablinger:2025joi,Ablinger:2024xtt}. In particular, we utilize the \texttt{libome}~\cite{Ablinger:2025joi,libome}  publicly available \texttt{C++} libraries for their numerical representation in terms of
precise local overlapping series expansion. 
For the 2-loop OMEs, we use the analytical expressions computed in Refs.~\cite{Buza:1995ie,Buza:1996wv}.  

\begin{table}[htp]
    \centering
    \vspace{5mm}
    \begin{tabular}{||c||r|r|r|r|r|r|r|r|}
    \hline \hline
    \multicolumn{9}{||c||}{} \\[-3mm]
    \multicolumn{9}{||c||}{$\, n_f = 3\ldots 5\,$,
        $\,\mu_{\rm f}^2 = 10^4 \mbox{ GeV}^2$} \\
    \multicolumn{9}{||c||}{} \\[-0.3cm]
    \hline \hline
    \multicolumn{9}{||c||}{} \\[-3mm]
	\multicolumn{1}{||c||}{$x$} &
    \multicolumn{1}{c|} {$xuv$} &
    \multicolumn{1}{c|} {$xdv$} &
    \multicolumn{1}{c|} {$xL_-$} &
    \multicolumn{1}{c|} {$xL_+$} &
    \multicolumn{1}{c|} {$xs_+$} &
    \multicolumn{1}{c|} {$xc_+$} &
    \multicolumn{1}{c|} {$xb_+$} &
    \multicolumn{1}{c||}{$xg$} \\[0.5mm]\hline \hline
\multicolumn{9}{||c||}{} \\[-3mm]
\multicolumn{9}{||c||}{$\mu_{\rm r}^2 = \ 1.0\mu_{\rm f}^2$} \\
\multicolumn{9}{||c||}{} \\[-0.3cm]
\hline \hline
  & & & & & & & & \\[-0.3cm]
 1e$-5$ &  3.017e$-3$ &  1.750e$-3$ &  1.281e$-4$ &  3.546e$+1$ &  1.706e$+1$ &  1.612e$+1$ &  1.315e$+1$ &  2.224e$+2$ \\
 1e$-4$ &  1.406e$-2$ &  8.228e$-3$ &  4.883e$-4$ &  1.561e$+1$ &  7.272e$+0$ &  6.785e$+0$ &  5.329e$+0$ &  8.859e$+1$ \\
 1e$-3$ &  6.082e$-2$ &  3.507e$-2$ &  1.763e$-3$ &  6.382e$+0$ &  2.779e$+0$ &  2.520e$+0$ &  1.851e$+0$ &  3.034e$+1$ \\
 1e$-2$ &  2.336e$-1$ &  1.307e$-1$ &  5.822e$-3$ &  2.267e$+0$ &  8.542e$-1$ &  7.045e$-1$ &  4.623e$-1$ &  7.786e$+0$ \\
 1e$-1$ &  5.485e$-1$ &  2.695e$-1$ &  9.996e$-3$ &  3.845e$-1$ &  1.125e$-1$ &  6.830e$-2$ &  3.790e$-2$ &  8.496e$-1$ \\
 3e$-1$ &  3.444e$-1$ &  1.276e$-1$ &  2.946e$-3$ &  3.457e$-2$ &  8.887e$-3$ &  3.966e$-3$ &  2.085e$-3$ &  7.870e$-2$ \\
 5e$-1$ &  1.179e$-1$ &  3.060e$-2$ &  3.653e$-4$ &  2.321e$-3$ &  5.681e$-4$ &  2.019e$-4$ &  1.138e$-4$ &  7.634e$-3$ \\
 7e$-1$ &  1.933e$-2$ &  2.965e$-3$ &  1.285e$-5$ &  5.242e$-5$ &  1.266e$-5$ &  3.402e$-6$ &  2.496e$-6$ &  3.709e$-4$ \\
 9e$-1$ &  3.315e$-4$ &  1.674e$-5$ &  8.104e$-9$ &  2.520e$-8$ &  6.644e$-9$ &  7.614e$-10$ &  1.432e$-9$ &  1.172e$-6$ \\
\hline \hline
\end{tabular}
\caption{aN$^3$LO evolution results from the \texttt{Candia-v2} code in the VFNS.}
\label{tab:n3lo_evolution_results}
\end{table}
\begin{table}[htp]
    \centering
    \vspace{5mm}
    \begin{tabular}{||c||r|r|r|r|r|r|r|r|r|r|r|}
    \hline \hline
    \multicolumn{12}{||c||}{} \\[-3mm]
    \multicolumn{12}{||c||}{$\, n_f = 3\ldots 5\,$,
        $\,\mu_{\rm f}^2 = 10^4 \mbox{ GeV}^2$} \\
    \multicolumn{12}{||c||}{} \\[-0.3cm]
    \hline \hline
    \multicolumn{12}{||c||}{} \\[-3mm]
	\multicolumn{1}{||c||}{$x$} &
    \multicolumn{1}{c|} {$g$} &
    \multicolumn{1}{c|} {$xu$} &
    \multicolumn{1}{c|} {$xd$} &
    \multicolumn{1}{c|} {$xs$} &
    \multicolumn{1}{c|} {$xc$} &
    \multicolumn{1}{c|} {$xb$} &
    \multicolumn{1}{c|} {$x\bar{u}$} &
    \multicolumn{1}{c|} {$x\bar{d}$} &
    \multicolumn{1}{c|} {$x\bar{s}$} &
    \multicolumn{1}{c|} {$x\bar{c}$} &
    \multicolumn{1}{c||}{$x\bar{b}$} \\[0.5mm]\hline \hline
\multicolumn{12}{||c||}{} \\[-3mm]
\multicolumn{12}{||c||}{$\mu_{\rm r}^2 = \ 1.0\mu_{\rm f}^2$} \\
\multicolumn{12}{||c||}{} \\[-0.3cm]
\hline \hline
  & & & & & & & & & & & \\[-0.3cm]
 1e$-5$ & $0.01$\% & $0.00$\% & $0.00$\% & $0.00$\% & $0.00$\% & $0.00$\% & $0.00$\% & $0.00$\% & $0.00$\% & $0.00$\% & $0.00$\% \\
 1e$-4$ & $0.00$\% & $0.00$\% & $0.00$\% & $0.00$\% & $0.00$\% & $0.00$\% & $0.00$\% & $0.00$\% & $0.00$\% & $0.00$\% & $0.00$\% \\
 1e$-3$ & $0.00$\% & $0.00$\% & $0.00$\% & $0.00$\% & $0.00$\% & $0.00$\% & $0.00$\% & $0.00$\% & $0.00$\% & $0.00$\% & $0.00$\% \\
 1e$-2$ & $0.00$\% & $0.00$\% & $0.00$\% & $0.00$\% & $0.00$\% & $0.00$\% & $0.00$\% & $0.00$\% & $0.00$\% & $0.00$\% & $0.00$\% \\
 1e$-1$ & $0.00$\% & $0.00$\% & $0.00$\% & $0.00$\% & $0.00$\% & $0.00$\% & $0.00$\% & $0.00$\% & $0.00$\% & $0.00$\% & $0.00$\% \\
 3e$-1$ & $0.00$\% & $0.00$\% & $0.00$\% & $0.00$\% & $0.01$\% & $0.00$\% & $0.00$\% & $0.00$\% & $0.00$\% & $0.01$\% & $0.00$\% \\
 5e$-1$ & $0.00$\% & $0.00$\% & $0.00$\% & $0.00$\% & $0.02$\% & $0.01$\% & $0.01$\% & $0.00$\% & $0.00$\% & $0.02$\% & $0.01$\% \\
 7e$-1$ & $0.00$\% & $0.00$\% & $0.00$\% & $0.00$\% & $0.06$\% & $0.03$\% & $0.04$\% & $0.00$\% & $0.00$\% & $0.06$\% & $0.03$\% \\
 9e$-1$ & $0.01$\% & $0.00$\% & $0.00$\% & $0.01$\% & $0.00$\% & $0.02$\% & $0.32$\% & $0.01$\% & $0.01$\% & $0.00$\% & $0.02$\% \\
\hline \hline
\end{tabular}
\caption{Percentage error between \texttt{Candia-v2} and \texttt{Hoppetv2}~\cite{Karlberg:2025hxk} evolution. The result from \texttt{Hoppet} is used as the reference.}
\label{tbl:hoppet-comparison}
\end{table}

In Table~\ref{tab:n3lo_evolution_results}, we report the output results of \texttt{Candia-v2} at aN$^3$LO at an evolution scale $\mu_F$ of 100 GeV, obtained with the approximate kernels discussed above.
The evolution is performed by using a grid split into three parts: a region from $x=10^{-5}$ to $x=0.1$ distributed logarithmically, a region from $x=0.1$ to $x=0.9$ distributed linearly, and a region from $x=0.9$ to $x=1.0$ distributed quadratically, with points packed towards large-$x$. The regions were filled with 200, 100, and 50 points respectively. We also used $s=20$ iterations in the NS and singlet sectors, and chose a truncation index of $\kappa=20$. Lastly, the convolutions were split into the same regions as above, each using 50 Gaussian points, with appropriate mappings (logarithmic, linear, and quadratic) in the corresponding regions. The results exhibit negligible variation with this parameter choice.

\begin{figure}[ht]
\centering
\includegraphics[width=0.89\linewidth]{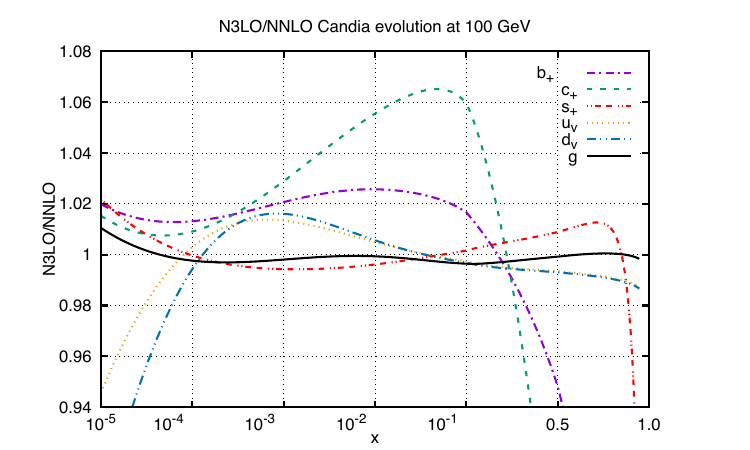}
\caption{N$^3$LO/NNLO PDF ratios in the VFNS scheme at $\mu_F=100~\unit{\GeV}$ with N$^3$LO matching conditions.}
\label{fig:ratios1}
\end{figure}

\begin{figure}
\centering
\includegraphics[width=0.89\linewidth]{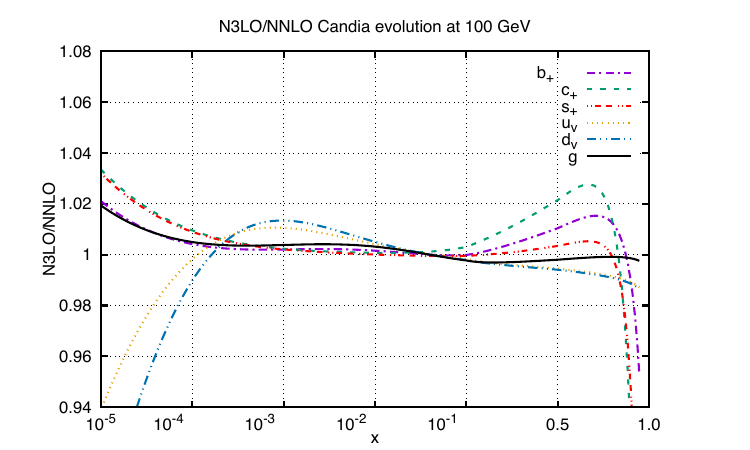}
\caption{Same as in Fig.~\ref{fig:ratios1}, but with NNLO matching conditions (with the $A^{(2)}$ OMEs only).}
\label{fig:ratios2}
\end{figure}

In Fig.~\ref{fig:ratios1}, we illustrate N$^3$LO/NNLO PDF ratios in the VFNS scheme with heavy-quark matching conditions at N$^3$LO, while in Fig.~\ref{fig:ratios2} the same PDF ratios are illustrated with matching conditions at NNLO, with the $A_{ij}^{(2)}$ OMEs only, as in Refs.~\cite{Karlberg:2026kte}. This allows for a separate assessment of impact from the splitting functions and OMEs. In particular, in Fig.~\ref{fig:ratios1} we observe an increase of approximately $6\%$ in the $c_+$ distribution in the $10^{-2}\lesssim x \lesssim 0.1$ range, and an increase of approximately $2.5\%$ in the $b_+$ combination in the $10^{-3}\lesssim x \lesssim 0.1$.  


In Table~\ref{tbl:hoppet-comparison}, we compare the results of \texttt{Candia-v2} to those from the recent \texttt{Hoppet-v2} computer code release, documented in Ref.~\cite{Karlberg:2026kte}. The $x$-space \texttt{Hoppet}'s algorithm is based on higher-order Runge-Kutta methods for both the $x$-integrations and $Q$ evolution. Table~\ref{tbl:hoppet-comparison} shows percentage errors, where the \texttt{Hoppet-v2} result is used as the reference. The differences between \texttt{Candia-v2} and  \texttt{Hoppet-v2} for the N$^3$LO evolution with approximate splitting functions are at the sub-percent level and below for all flavors. 
We show individual PDF distributions rather than $\pm$ combinations to avoid the introduction of numerical errors due to large cancellations between these PDFs at low and large $x$ as well as in the computation of the percentage errors between \texttt{Candia-v2} and \texttt{Hoppet-v2} in the same kinematic ranges. 

\section{Numerical Optimizations}
\label{optmiz}

In this section, we discuss key improvements and optimizations in \texttt{Candia-v2}’s numerical implementation, leveraging the algorithm’s recursive structure. 

In both the singlet and non-singlet sectors, the ``outer'' iteration over the index $s$ only requires information from the previous value of $s$. As a result, in the code a large amount of memory can be saved by allocating only two sets of arrays (which themselves may have several dimensions). 
This issue was not addressed in the original \texttt{C} implementation of \texttt{Candia} and can be particularly severe at N$^3$LO, where the presence of an additional dimension leads to exponentially growing memory requirements.

Moreover, all convolutions are performed using Gauss–Legendre quadrature with a fixed set of weights and abscissae. Consequently, the splitting functions and operator matrix elements need to be evaluated only at a finite, fixed set of points—the quadrature abscissae. This allows the splitting functions evaluated at these abscissae to be cached so that, in subsequent convolutions, the cached values can be reused instead of repeatedly recomputing the splitting functions, which may become costly at higher perturbative orders. This optimization achieves a 20–25\% average speedup for the exact solution by efficiently handling the exponential increase in splitting function and operator matrix element evaluations. However, improvements for the truncated solution are marginal, and the method incurs a higher memory cost that scales with grid size.

The final major optimization exploits the fact that the non-singlet PDFs are completely decoupled. This property allows the evolution of the non-singlet PDFs to be parallelized by assigning them to separate threads. Apart from the overhead associated with thread creation and management, this strategy can in principle yield an approximately 
$N$-fold speedup in the non-singlet sector when using $N$ threads. 
Further, while the singlet sector evolves a set of coupled PDFs, meaning they cannot be assigned to separate threads, we can still split up the convolution into separate pieces and perform them concurrently. This is possible due to the fact that, as mentioned previously, the convolutions require information only from the previous iteration.
After implementing these optimizations, \texttt{Candia-v2} exhibits significantly improved performance relative to \texttt{Candia} at higher perturbative orders. The initial N$^3$LO evolution code, structured as an extension of the \texttt{Candia} code, exhibited execution times surpassing 20 minutes for high-precision runs. Our optimized implementation reduces this latency to less than one minute while simultaneously achieving enhanced precision.

Further improvements to the \texttt{Candia-v2} computer code are currently underway and are expected to enhance both numerical precision and CPU performance. These developments will be presented in a forthcoming publication, which will represent the official new release of \texttt{Candia-v2}.

\section{Conclusions}
\label{conclusions}

In this work, we have extended the $x$-space \texttt{Candia} algorithm for solving the DGLAP evolution equations for unpolarized collinear PDFs to N$^3$LO in pQCD. The approach, based on logarithmic expansions of DGLAP solution, is fully general and can be carried out to arbitrary perturbative order. 
Furthermore, it provides insight into the analytical structure of DGLAP solutions in 
$x$-space, owing to the simplicity of the logarithmic expansions from which recursion relations are obtained.
The algorithm is implemented in a new version of the \texttt{Candia} computer code~\cite{candiav1} named \texttt{Candia-v2}, which is publicly available on GitHub~\cite{candiav2}.
In the non-singlet sector, we have derived an exact power-series expansion that is equivalent to the exact N$^3$LO solution to the DGLAP equations. For both the singlet and non-singlet sectors, we obtained extended recursion relations for the truncated solutions to the N$^3$LO DGLAP equations at arbitrary perturbative order.  
In addition, we have presented benchmark results for the aN$^3$LO evolution using the most recent approximations to the N$^3$LO splitting functions, and a comparison with the new release of the $x$-space \texttt{Hoppet-v2} computer code~\cite{Karlberg:2026kte}.  
Further developments in \texttt{Candia-v2} are in progress which will improve its current precision and performance. They will be presented in a forthcoming publication for the official new release of \texttt{Candia-v2}. The \texttt{Candia} algorithm offers a flexible and efficient framework for high-precision parton evolution in QCD at higher perturbative orders, and provides insight into the analytical structure of DGLAP solutions in the $x$-space.

\begin{acknowledgments}
M.G. and C.H. are partially supported by the National Science Foundation under Grant No.~PHY-2412071. This work used the high-performance computing resource from the KSU HPC clusters.
\end{acknowledgments}

\bibliographystyle{utphys}

\providecommand{\href}[2]{#2}
\begingroup\raggedright

\endgroup

\end{document}